\documentclass[conference]{IEEEtran}
\IEEEoverridecommandlockouts
\usepackage{cite}
\usepackage{amsmath,amssymb,amsfonts}
\usepackage{graphicx}
\usepackage{textcomp}
\usepackage{xcolor}
\usepackage{caption}
\usepackage{braket}
\usepackage{algpseudocode}
\usepackage{algorithm}
\usepackage{booktabs}
\usepackage{multirow}
\usepackage{verbatim}
\usepackage{caption}
\usepackage{subcaption}

\usepackage{tikz}
\usetikzlibrary{shapes.geometric, arrows, backgrounds, fit}

\tikzstyle{block} = [rectangle, fill=yellow!20, draw=yellow, text centered, rounded corners, minimum height=2em, minimum width=5em, text width=7em]
\tikzstyle{block1} = [rectangle, fill=yellow!20, draw=yellow, text centered, rounded corners, minimum height=2em, minimum width=5em, text width=10em]
\tikzstyle{blockb} = [rectangle, fill=cyan!10, draw=cyan, text centered, rounded corners, minimum height=2em, minimum width=5em, text width=7em]
\tikzstyle{blockf} = [rectangle, fill=cyan!10, draw=cyan, text centered, rounded corners, minimum height=2em, minimum width=5em, text width=12em]
\tikzstyle{decision} = [diamond, fill=yellow!20, draw=yellow, text centered, aspect=2, minimum height=2em, minimum width=4em]
\tikzstyle{background} = [rectangle, fill=gray!20, inner sep=0.2cm, rounded corners=5mm, very thick, dashed, draw=gray!70]
\tikzstyle{background0} = [rectangle, fill=gray!10, inner sep=0.5cm, rounded corners=5mm]

\tikzstyle{arrow} = [thick,->,>=stealth]

\def\BibTeX{{\rm B\kern-.05em{\sc i\kern-.025em b}\kern-.08em
    T\kern-.1667em\lower.7ex\hbox{E}\kern-.125emX}}

\newcommand{\YC}[1]{\textcolor{blue}{YC: #1}}

\newcommand{\Yidong}[1]{\textcolor{magenta}{Yidong: #1}}

\begin{document}

\title{ECDQC: Efficient Compilation for Distributed Quantum Computing with Linear Layout\thanks{The views expressed in this article are those of the authors and do not represent the views of Wells Fargo. This article is for informational purposes only. Nothing contained in this article should be construed as investment advice. Wells Fargo makes no express or implied warranties and expressly disclaims all legal, tax, and accounting implications related to this article.}}

\author{\IEEEauthorblockN{
Kecheng Liu*\textsuperscript{1}  \ \
Yidong Zhou*\textsuperscript{1}  \ \
Haochen Luo\textsuperscript{2}  \ \
Lingjun Xiong\textsuperscript{1}  \ \
Yuchen Zhu\textsuperscript{1}  \ \
Eilis Casey\textsuperscript{1}\\
Jinglei Cheng\textsuperscript{3}    \ \
Samuel Yen-Chi Chen\textsuperscript{4}  \ \
Zhiding Liang\textsuperscript{1}}
\IEEEauthorblockA{
\textsuperscript{1}Rensselaer Polytechnic Institute
\textsuperscript{2}Cornell University
\textsuperscript{3}University of Pittsburgh
\textsuperscript{4}Wells Fargo\\
*These authors contributed to the work equally and should be regarded as co-first authors.\\
Corresponding author: liangz9@rpi.edu
\vspace{-0.15in}}
}

\maketitle

\begin{abstract}
In this paper, we propose an efficient compilation method for distributed quantum computing (DQC) using the Linear Nearest Neighbor (LNN) architecture. By exploiting the LNN topology's symmetry, we optimize quantum circuit compilation for High Local Connectivity, Sparse Full Connectivity (HLC-SFC) algorithms like Quantum Approximate Optimization Algorithm (QAOA) and Quantum Fourier Transform (QFT). We also utilize dangling qubits to minimize non-local interactions and reduce SWAP gates. Our approach significantly decreases compilation time, gate count, and circuit depth, improving scalability and robustness for large-scale quantum computations.
\end{abstract}


\section{Introduction}
Quantum computation promises faster solutions than classical algorithms in fields like factorization \cite{Shor}, finance~\cite{herman2022survey}, and chemistry~\cite{cao2019quantum}. As quantum computing progresses from the Noisy Intermediate-Scale Quantum (NISQ) era \cite{preskillQuantumComputingNISQ2018,nash2020quantum} toward Fault-Tolerant Quantum Computing (FTQC) \cite{shor1996fault,gaitan2008quantum}, many applications demand more resources than a single NISQ device can provide. Distributed quantum computing (DQC) offers a way to scale quantum systems by utilizing multiple processors \cite{burt2024generalisedcircuitpartitioningdistributed, DQC-survey}.

In the framework of DQC, each quantum computer functions as an independent quantum processing unit (QPU), with quantum teleportation facilitating the communication between different QPUs~\cite{burt2024generalisedcircuitpartitioningdistributed}. 
Quantum teleportation, implemented using non-local two-qubit gates, is highly susceptible to system coherence times and noise, which represent two of the primary challenges in DQC systems. Additionally, conventional compilation on Heavy-hex topology introduces excess number of swap gates, leading to increased compilation overhead, deeper circuits, and higher errors~\cite{crosstalk}.

\begin{figure}[h]
    \centering
    \includegraphics[width=0.7\linewidth]{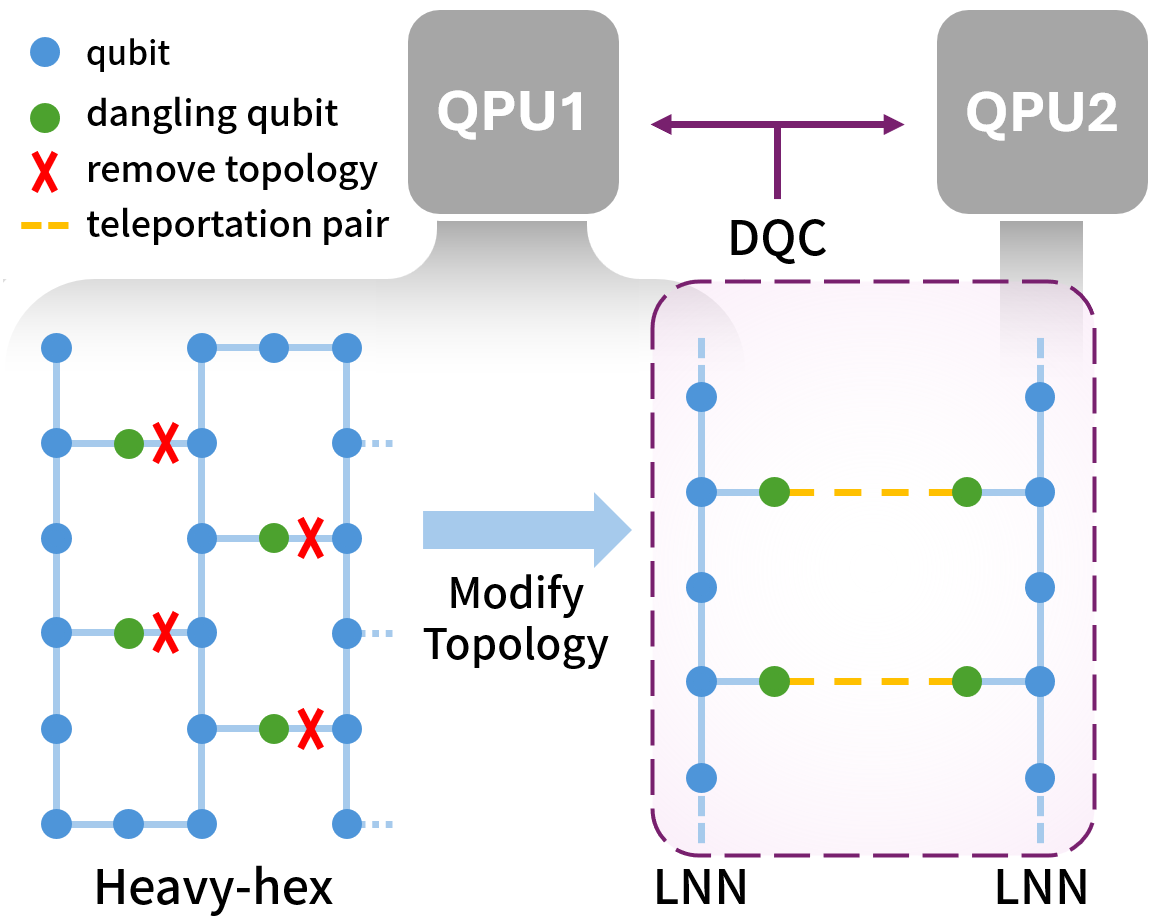}
    \caption{Proposed chip-to-chip distributed quantum system by modifying the Heavy-hex topology to LNN topology and taking the dangling qubits as teleportation pair.}
    \label{fig:overview}
\end{figure}

To address these challenges, as illustrated in Figure~\ref{fig:overview}, we restructure the Heavy-hex topology into a Linear Nearest Neighbor (LNN) topology, removing specific qubit connections to reduce compilation overhead~\cite{zhu2024coqablazingfastcompiler}. This transformation results in dangling qubits within the LNN structure, which can serve as teleportation qubits, allowing for the application of non-local gates to facilitate communication between QPUs.

For quantum chips with the same Heavy-hex topology, the topological transformation will result in the same LNN structure with dangling qubits. 
This symmetry structure allows dangling qubits to communicate with each other one-to-one, thus deciding the structure of the DQC system. 
This structure also allows DQC to use LNN compilation directly, reduce noise and circuit depth, which makes DQC system more reliable and practical.

\section{Background}

\subsection{Quantum Teleportation}
Quantum teleportation facilitates the transfer of a quantum state from one location to another without directly transmitting qubits. This process involves three concepts: quantum states, quantum entanglement, and classical communication \cite{Bennett_2014}.


Quantum states represent the complete information of a quantum system, often in the format of wave functions or state vectors. 
Quantum entanglement describes the correlated states of multiple systems. 
Classical communication is the key for transmitting measurement results in quantum teleportation. 
As entanglement is an important component of DQC, many protocols focus on optimizing the entanglement costs of gate or state teleportation. 
Within the framework of DQC, quantum teleportation is classified into two distinct categories: state teleportation and gate teleportation \cite{burt2024generalisedcircuitpartitioningdistributed}.

Quantum state is a complete description of a quantum physical system, typically represented as rays in Hilbert space. We can express a qubit in a superposition state as $\ket{\phi}=\alpha\ket{0}+\beta\ket{1}$, where $\alpha$, $\beta$ are complex numbers that satisfy $|\alpha|^{2}+|\beta|^{2}=1$. We consider any bipartite pure state $\ket{\psi}_{AB}$, where $A$, $B$ represent two Hilbert spaces $\mathcal{H}_{A}$, $\mathcal{H}_{B}$. $\ket{\psi}_{AB}$ is entangled if its Schmidt number is greater than one. A typical entangled state is the Bell state $\ket{\Phi^{+}}=\frac{1}{\sqrt{2}}(\ket{00}+\ket{11})$. The following presents a specific example of quantum teleportation based on this Bell state \cite{Nielsen_Chuang_2010, Bouwmeester_1997}.

Suppose Alice wants to teleport her qubit $\ket{\phi}_{A_{1}}=\alpha\ket{0}_{A_{1}}+\beta\ket{1}_{A_{1}}$ to Bob, who has an entangled pair shared with Alice, represented as $\ket{\Phi^{+}}_{A_{2}B}=\frac{1}{\sqrt{2}}(\ket{0}_{A_{2}}\ket{0}_{B}+\ket{1}_{A_{2}}\ket{1}_{B})$. The combined state of Alice's qubit and the entangled pair is $\ket{\psi}_{A_{1}A_{2}B}=\ket{\phi}_{A_{1}}\otimes\ket{\Phi^{+}}_{A_{2}B}$. Then Alice performs a Bell measurement on her qubit and her part of the entangled state, yielding one of the four Bell states $\ket{\Phi^{\pm}}$ and $\ket{\Psi^{\pm}}$ :
\begin{equation}
    \begin{aligned}
        \ket{\psi}_{A_{1}A_{2}B}
        =&\frac{1}{2}\ket{\Phi^{+}}_{A_{1}A_{2}}(\alpha\ket{0}_{B}+\beta\ket{1}_{B})\\
        +&\frac{1}{2}\ket{\Phi^{-}}_{A_{1}A_{2}}(\alpha\ket{0}_{B}-\beta\ket{1}_{B})\\
        +&\frac{1}{2}\ket{\Psi^{+}}_{A_{1}A_{2}}(\beta\ket{0}_{B}+\alpha\ket{1}_{B})\\
        +&\frac{1}{2}\ket{\Psi^{-}}_{A_{1}A_{2}}(\alpha\ket{1}_{B}-\beta\ket{0}_{B}).
    \end{aligned}
\end{equation}

After measurement, Alice sends the result of her measurement to Bob. Depending on the measurement outcome, Bob applies a corresponding unitary operation to his part of the entangled state. After this operation, Bob's qubit will be in the state $\ket{\phi}$, completing the teleportation process \cite{RevModPhys.81.865}.
Within the framework of DQC, quantum teleportation is classified into two distinct categories: state teleportation and gate teleportation \cite{burt2024generalisedcircuitpartitioningdistributed}.

\subsubsection{State teleportation}
Quantum state teleportation is the standard teleportation procedure. 
The entity being transmitted is a quantum state, incurred at the cost of consuming one entangled bit (e-bit) \cite{PhysRevLett.70.1895}. 
This leads us to utilize state teleportation to cover a non-local operation by transferring the state of a qubit to its partner's QPU and subsequently executing the operation locally, also at the expense of consuming one e-bit.

\subsubsection{Gate teleportation}

Quantum gate teleportation is an unconventional method that uses a single e-bit to execute a two-qubit controlled unitary operation without transferring qubits between QPUs \cite{PhysRevLett.123.070505}. 
It entangles the control qubit with a communication qubit in a separate QPU, allowing identical control over two-qubit gates on other qubits. 
The communication qubit is measured after the gate operation, and its result is used to correct the control qubit’s state. 
Unlike state teleportation, both qubits remain in their original QPUs\cite{chou2018deterministic}.

Recent work shows that larger sequences of two-qubit operations can sometimes be teleported without additional cost \cite{PhysRevA.62.052317, huang2004experimental}. 
For extended operations, when gates share a control qubit and target qubits are on the same QPU, the communication qubit’s measurement can be delayed until all gates are executed, reducing e-bit consumption \cite{Wu2023entanglement}.

\subsection{Chip to Chip Distribution}
DQC chip-to-chip distributed refers to the ability to perform quantum computing across multiple interconnected quantum chips. This approach enables various quantum processors to collaborate effectively, addressing complex quantum computing challenges together \cite{9268630}. This approach is particularly well-suited for applications that demand a substantial number of quantum bits and involve complex computations, such as quantum simulation\cite{zhang2022unbiased, granet2024hamiltonian}, optimization problems\cite{moll2018quantum,liang2024graph}, and quantum machine learning (QML) \cite{biamonte2017quantum, wang2024separable}.


\subsubsection{Architecture}

In this architecture, each quantum chip has its own qubits and gates. 
Through teleportation and manipulation of quantum states, multiple chips collaborate in computation. 
For example, the dangling qubits in the linear compiler enable state transmission between chips.
This approach requires efficient state transmission and quantum error correction (QEC) \cite{sivak2023real,xu2022distributed } to preserve information integrity and reduce noise\cite{llewellyn2020chip}.

\subsubsection{Implementation}
Quantum teleportation is essential for DQC, enabling quantum state transmission between remote processors without moving qubits. 
It relies on entangled pairs shared across chips to securely transfer quantum information. 
The process begins with a Bell-state measurement, generating classical information sent to the receiver chip, which reconstructs the original state using quantum gates. 


Quantum teleportation facilitates indirect state transfer via entanglement and classical communication, reducing exposure to environmental disturbances and relying on pre-shared entangled states that offer robustness against local noise. This approach enhances scalability and fault tolerance by mitigating noise and decoherence, thereby laying the groundwork for large-scale quantum networks and advancing applications \cite{inc2024distributedquantumcomputingsilicon}.

\section{Motivation}
\label{section_motivation}
The need for efficient quantum circuit compilation in DQC arises from the scalability challenges inherent in large quantum systems. As the qubit count in circuits of quantum algorithms characteristic of the NISQ and FTQC increases, linear compilers become impractical due to increased gate depth, qubit swaps, and inter-chip communication overhead. However, the linear compiler’s symmetric structure offers a unique insight: its perfect symmetry in qubit distribution makes it a natural fit for adaptation into a distributed compilation methodology.

The structure of the linear compiler inherently provides a balanced distribution of qubits, which lends itself well to a distributed architecture. This symmetry can be leveraged to distribute qubits across multiple QPUs while maintaining efficient gate execution and minimal overhead. Thus, the linear compiler's structure is the foundation for our distributed optimization methodology.

Moreover, through the analysis of the linear compiler, we notice that it frequently leaves some qubits ``dangling'' at the ends of qubit chains. These dangling qubits are ideal candidates for inserting non-local gates, as they often require interactions with qubits on different QPUs. By recognizing and optimizing the placement of these dangling qubits, we can further reduce the number of SWAP gates and improve the overall performance of distributed quantum circuits. This insight is crucial to our distributed compiler design, allowing us to efficiently manage non-local operations without incurring excessive overhead.

Chip-to-chip DQC addresses these issues by optimizing qubit placement, minimizing cross-QPU communication, and employing techniques like entanglement swapping and data-qubit swapping to execute operations efficiently.




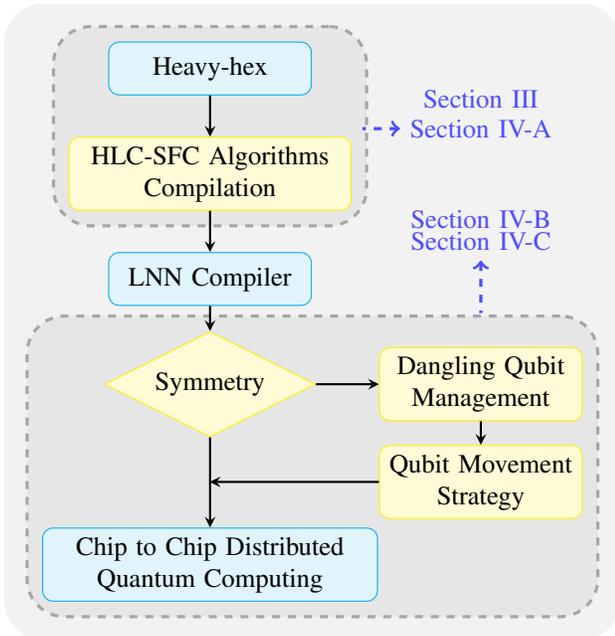
\begin{figure}[t]
    \centering
    
\begin{tikzpicture}[node distance=1.4cm]

    \node (heavyhex) [blockb, xshift=1cm] {Heavy-hex};
    \node (variational) [block1, below of=heavyhex] {HLC-SFC Algorithms Compilation};
    \node (lnncompiler) [blockb, below of=variational] {LNN Compiler};
    \node (symmetry) [decision, below of=lnncompiler] {Symmetry};
    \node (dangling) [block, right of=symmetry, xshift=2.2cm] {Dangling Qubit Management};
    \node (qubitmove) [block, below of=dangling, yshift=0.1cm] {Qubit Movement Strategy};
    \node (final) [blockf, below of=symmetry, yshift=-1cm] {Chip to Chip Distributed Quantum Computing};

    \draw [arrow] (heavyhex) -- (variational);
    \draw [arrow] (variational) -- (lnncompiler);
    \draw [arrow] (lnncompiler) -- (symmetry);
    \draw [arrow] (symmetry) -- (final);
    \draw [arrow] (symmetry) -- (dangling);
    \draw [arrow] (dangling) -- (qubitmove);
    \draw [arrow] (qubitmove.west) -- +(left:2.25cm);

    \node[align=left, text=blue!70, thick, right of=variational, xshift=2.2cm, yshift=1cm](section3) {Section \ref{section_motivation}};
    \node[align=left, text=blue!70, thick, below of=section3, yshift=1cm](section41) {Section \ref{section_LNN}};
    \node[align=left, text=blue!70, thick, right of=lnncompiler, xshift=2.2cm, yshift=0.8cm](section42) {Section \ref{section_qm}};
    \node[align=left, text=blue!70, thick, below of=section42, yshift=1.1cm](section43) {Section \ref{section_ms}};

    \draw [dashed, very thick, blue!70, <-] (section41.west) -- +(left:0.5cm);
    \draw [dashed, very thick, blue!70, <-] (section43.south) -- +(south:0.7cm);

    \begin{pgfonlayer}{background}
    \node [background0, fit=(heavyhex)(final)(dangling)(qubitmove)] {};
    \node [background, fit=(heavyhex)(variational)] {};
    \node [background, fit=(symmetry)(dangling)(qubitmove)(final)] {};
    \end{pgfonlayer}

\end{tikzpicture}
\vspace{0.5em}
    \caption{\textbf{Overview of the Chip-to-Chip DQC Compilation Process.} The process begins with a heavy-hex architecture and compiles HLC-SFC quantum algorithms, such as QAOA and QFT, using an LNN compiler. The compiler leverages symmetry to optimize the structure, managing dangling qubits and employing a qubit movement strategy. This leads to an efficient chip-to-chip DQC system with minimized non-local interactions and enhanced scalability.}
    \label{fig:workflow-label}
\vspace{-3mm}
\end{figure}

\section{Methodology}
The scalability challenges in DQC arise as quantum systems grow larger, with increasing qubit counts leading to deeper circuits, more qubit swaps, and greater communication overhead. However, the symmetric structure of the linear compiler offers a solution: its balanced qubit distribution naturally fits a distributed compilation approach, reducing non-local interactions and SWAP gates.

By leveraging this symmetry, qubits can be efficiently distributed across multiple QPUs while maintaining minimal overhead. Additionally, the frequent presence of "dangling" qubits in the linear compiler makes them ideal for inserting non-local gates, further reducing SWAP gates and improving overall circuit performance in distributed systems. This insight informs our design for an optimized distributed compiler that effectively handles non-local operations.

\subsection{Linear Structure with Dangling Point}
\label{section_LNN}
The linear compiler operates within a LNN architecture, where qubits are arranged in a sequence and only adjacent qubits can interact directly. This structure is simple, making it effective for various quantum algorithms, including the Quantum Approximate Optimization Algorithm (QAOA)\cite{farhi2014quantum} in NISQ and the Quantum Fourier Transform (QFT)\cite{weinstein2001implementation} in FTQC. Both QAOA and QFT share a common trait: they exhibit High Local Connectivity (HLC) with Sparse Full Connectivity (SFC). This means that most interactions in these algorithms occur between nearby qubits, while only a few require non-local connections. These characteristics reduce the demand for long-range interactions, making them well-suited for LNN architectures where minimizing non-local SWAP operations is critical for optimizing performance. We refer to these as HLC-SFC algorithms, as they balance strong local interactions with minimal long-distance connectivity. The key advantage of LNN is that it minimizes the complexity of the qubit interaction graph, allowing for easier compilation and fewer SWAP gates, which are typically required in more complex architectures like 2D grids or heavy-hex lattices.


In the linear compiler, qubits are often organized into lines with "dangling points" — qubits that only have one neighboring qubit. These dangling qubits are strategically used to reduce the number of non-local interactions and SWAP gates. By placing less critical qubits in these positions, the compiler ensures that the overall structure remains efficient. Additionally, the dangling points can serve as storage for qubits that are not immediately needed in the computation, reducing unnecessary movements.

When compiling a quantum circuit, the linear compiler begins by assigning qubits to specific positions within the linear structure, ensuring that interactions between qubits are as local as possible. If two qubits need to interact but are not adjacent, the compiler uses a series of SWAP operations to bring them closer together. However, by leveraging the dangling points, the compiler minimizes the number of these costly operations, thereby reducing the gate count and improving run-time efficiency. This strategy makes the linear compiler particularly well-suited for HLC-SFC algorithms, which do not require fully connected qubit architectures. The simplicity of the LNN model allows for a streamlined compilation process that efficiently maps quantum circuits onto hardware with minimal overhead. During quantum circuit compilation, the linear compiler assigns qubits to positions within the linear structure, keeping interactions as local as possible\cite{zhu2024coqablazingfastcompiler}. 


If non-adjacent qubits require interaction, SWAP operations are used to bring them closer physically.
Leveraging dangling points minimizes these costly operations, reducing gate count and improving runtime efficiency. 
This makes the linear compiler ideal for HLC-SFC algorithms, which don’t need fully connected qubit architectures, ensuring a more efficient process for hardware mapping.

\subsection{Dangling Qubits Management}
\label{section_qm}
In the linear compiler, the special structure gives out a perfect and natural inspiration for using DQC. 
Especially when it comes to the dangling qubits, we notice that no matter how the non-local gate moves through the dangling qubits pair, the topological structure will remain the same and won’t influence the performance of DQC. 
This unique aspect of dangling qubits in an LNN topology provides an excellent opportunity to optimize quantum operations by strategically utilizing these qubits to minimize the number of non-local operations and SWAP gates required.

Dangling qubits are typically located at the ends or breakpoints of qubit chains within the linear architecture. Their strategic position makes them less entangled with the main computational workflow, which allows them to serve as ideal candidates for temporary storage or for the execution of specific quantum gates that do not interfere with the primary computational paths. This usage reduces the need for qubit relocations, which are both time-consuming and error-prone, thereby enhancing the overall efficiency and reliability of the quantum computation.

This reduces the need for time-consuming qubit re-allocations, improving computational efficiency and reliability.

The management of dangling qubits is crucial in reducing the compilation complexity. By aligning qubits in such a way that interactions predominantly occur between adjacent qubits, the linear compiler significantly limits the necessity for long-range interactions, which are traditionally costly in terms of both quantum coherence and operational overhead. When non-local interactions are necessary, dangling qubits can be employed to facilitate these interactions through quantum teleportation or other entanglement-based operations, thus bypassing the need for extensive qubit movements. The optimization of dangling qubits also involves the dynamic reconfiguration of their roles depending on the computational phase. During phases where fewer qubits are required, these dangling qubits can act as passive elements, effectively out of the way of main computational tasks. However, during more complex operations, they can be activated to perform critical tasks such as entanglement generation or as intermediate nodes in multi-qubit gates.


\subsection{Qubit Movement Strategy}
\label{section_ms}
The qubit movement strategy within the framework of the linear compiler focuses on minimizing the physical movement of qubits across the quantum processor, thereby reducing the complexity and duration of computations. 
Qubit movement is strategically minimized by leveraging the linear structure of the LNN topology.
This structure inherently limits the necessity for qubit re-allocations, as qubits are primarily arranged in a straight line and only interact with their nearest neighbors. 
This configuration simplifies the control logic required for qubit interactions and significantly reduces the potential for operational errors associated with qubit movement. 
By prioritizing the movement of qubits that are closer to their interaction targets, the compiler effectively decreases the total number of SWAP operations needed, which in turn reduces the overall gate count and enhances the circuit’s performance.

The effectiveness of this qubit movement strategy is underscored by our comparative results with the SABRE compiler~\cite{li2019tackling}. In benchmark tests, our linear compiler demonstrated a significant reduction in gate count and circuit depth, leading to improved run-time performance.




\section{Evaluation}
\label{section_evaluation}

\subsection{Experiment Setting}
We evaluate our linear DQC method through two experiments, both targeting IBM’s quantum architecture. The first compares our linear compiler \cite{zhu2024coqablazingfastcompiler} with the SABRE compiler, focusing on gate count and circuit depth. The second experiment examines our non-local gate placement strategy, demonstrating the advantages of placing gates on dangling qubits to reduce SWAP operations and improve circuit performance. We evaluate our linear DQC method through two experiments. The first compares our linear compiler \cite{zhu2024coqablazingfastcompiler} with the SABRE compiler, assessing gate count and circuit depth. SABRE experiments are run on an IBM quantum machine with architecture constraints, while the linear compiler is tested on a local simulator. The second experiment is focused on our non-local gate placement strategy, highlighting the advantages of placing gates on dangling qubits. All distributed compiler tests are conducted on the local simulator. 

\subsection{Experiment 1: Comparison Between SABRE and Linear Compiler}
This experiment is aimed to compare the performance of the traditional SABRE compiler with our linear compiler. 
Key performance metrics including gate count and circuit depth are adopted for evaluation across various QAOA and QFT circuits as the run-time advantage of our linear compiler has already been demonstrated in previous work~\cite{zhu2024coqablazingfastcompiler, jin2021structured}.

The results, as summarized in Table~\ref{tab:sabre_vs_linear_qaoa} and Table~\ref{tab:sabre_vs_linear_qft}, show clear advantages for the linear compiler. 
For instance, in a 100-qubit QFT circuit, the linear compiler reduces the gate count by 48\%, from 148,398 (SABRE) to 76,795 gates, and decrease the circuit depth by 25\%.

\begin{table}[h]
    \centering
    \caption{Comparison of Gate Count and Circuit Depth between SABRE and Linear Compiler for QAOA Circuits.}
    \begin{tabular}{cccccc}
    \toprule
         \multirow{2}{*}{Qubit Count} & \multicolumn{2}{c}{Gate Count} & \multicolumn{2}{c}{Circuit Depth} \\
         & SABRE & Linear & SABRE & Linear \\
    \midrule
         10  & 408 & 175 & 200 & 83 \\
         20  & 1631 & 730 & 625 & 193 \\
         50  & 9895 & 4675 & 2136 & 523 \\
         80  & 26149 & 12040 & 3819 & 853 \\
         100  & 40837 & 18850 & 5031 & 1073 \\
         120  & 59923 & 27180  & 6097 & 1293 \\
    \bottomrule
    \end{tabular}
    \label{tab:sabre_vs_linear_qaoa}
\end{table}

\begin{table}[h]
    \centering
    \caption{Comparison of Gate Count and Circuit Depth between SABRE and Linear Compiler for QFT Circuits.}
    \begin{tabular}{cccccc}
    \toprule
         \multirow{2}{*}{Qubit Count} & \multicolumn{2}{c}{Gate Count} & \multicolumn{2}{c}{Circuit Depth} \\
         & SABRE & Linear & SABRE & Linear \\
    \midrule
         10   & 109    & 107     & 81    & 59  \\
         20   & 523    & 487     & 407   & 126 \\
         50   & 4176   & 3076    & 2577  & 376 \\
         80   & 11676  & 8164    & 6472  & 617 \\
         100  & 20664  & 12854   & 10849 & 784 \\
         120  & 30479  & 18671   & 15959 & 936 \\
    \bottomrule
    \end{tabular}
    \label{tab:sabre_vs_linear_qft}
\end{table}

These improvements stem from LNN's efficient qubit placement, minimizing SWAP gates for non-local interactions. SABRE excels in smaller circuits but faces challenges with larger ones due to hardware topology constraints, leading to higher gate counts and deeper circuits.

\subsection{Experiment 2: Non-Local Gate Placement with Dangling Qubits}

In this experiment, we introduce a new metric: cross-group SWAP, which refers to the count of qubits that need to traverse a non-local gate via a SWAP operation during compilation. This metric is particularly important in DQC because it directly impacts both the complexity and the fidelity of the circuit. The significance of reducing cross-group SWAP gates, and how it enhances the performance of distributed quantum systems, are further discussed in Section\ref{section_conclusion}.
Here we introduce a new metric: cross-group SWAP, measuring qubits that traverse non-local gates via SWAP operations during compilation. This metric is crucial in DQC, affecting both circuit complexity and fidelity, with further discussion in Section \ref{section_conclusion}.

\begin{figure}[h]
    \centering
    \begin{subfigure}[t!]{0.45\linewidth}
        \centering
        \includegraphics[width=\linewidth]{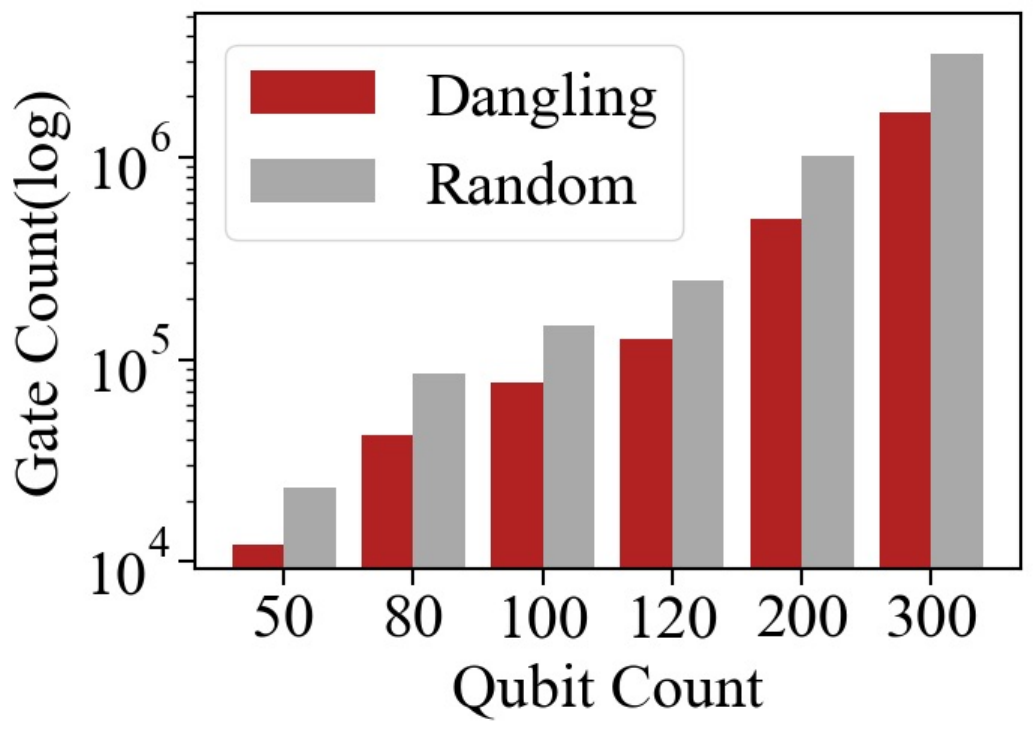}    
        \label{fig:qftgate}
        \caption{Gate count of QFT} 
    \end{subfigure}
    \begin{subfigure}[t!]{0.45\linewidth}  
        \centering 
        \includegraphics[width=\linewidth]{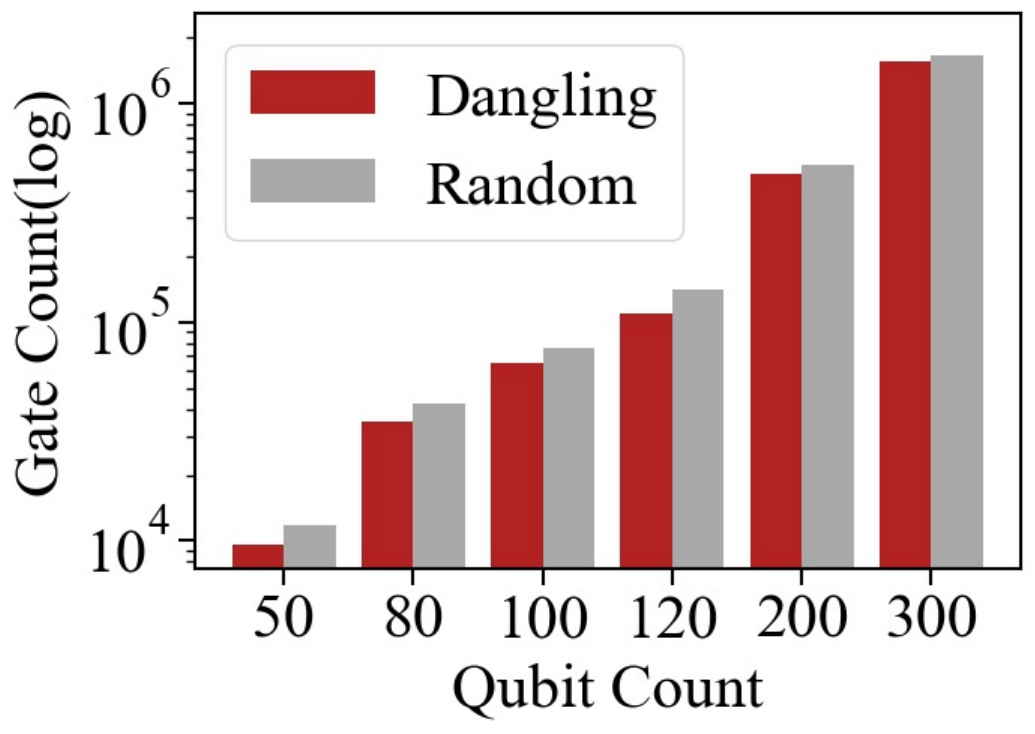}  
        \label{fig:qaoagate}
        \caption{Gate count of QAOA}
    \end{subfigure}
    \caption{Gate Count Comparison between Random and Dangling Qubit Placement for QFT and QAOA Circuits.} 
    \label{fig:random_vs_dangling_gate_count}
\end{figure}

\begin{figure}[h]
     \centering
     \begin{subfigure}[b]{0.22\textwidth}  
        \includegraphics[width=\linewidth]{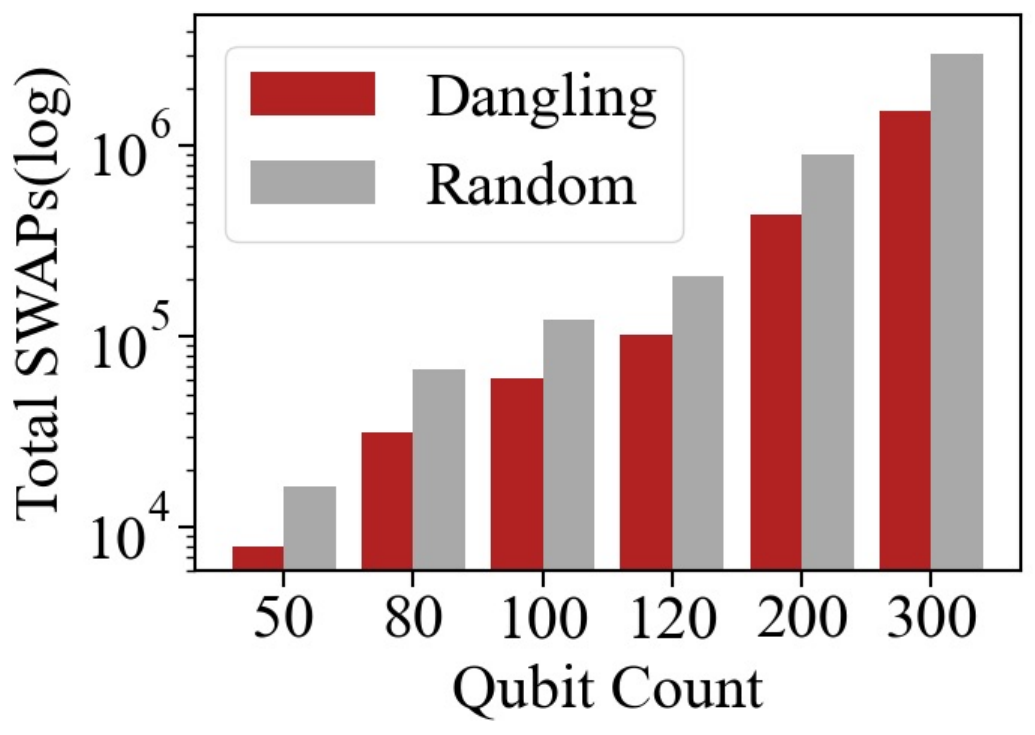}
        \caption{QFT SWAP}
        \label{subfig:qftswap}
     \end{subfigure}
     \hfill  
     \begin{subfigure}[b]{0.22\textwidth}  
        \includegraphics[width=\linewidth]{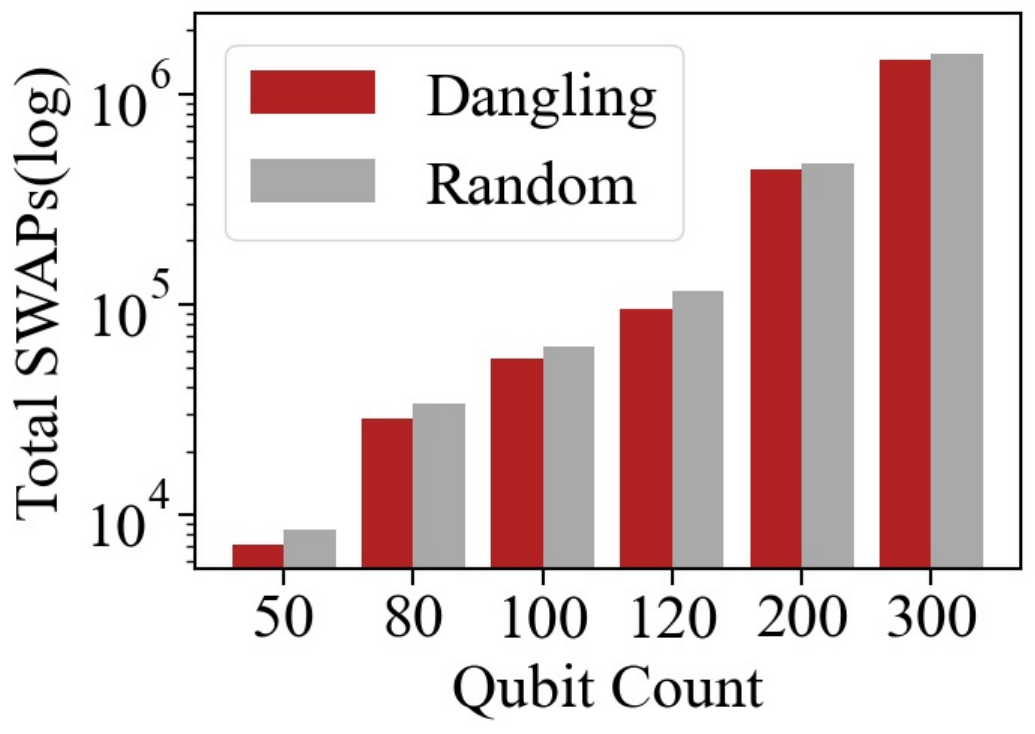}
        \caption{QAOA SWAP}
        \label{subfig:qaoaswap}
     \end{subfigure}
     \vskip\baselineskip  
     \begin{subfigure}[b]{0.22\textwidth}  
         \includegraphics[width=\linewidth]{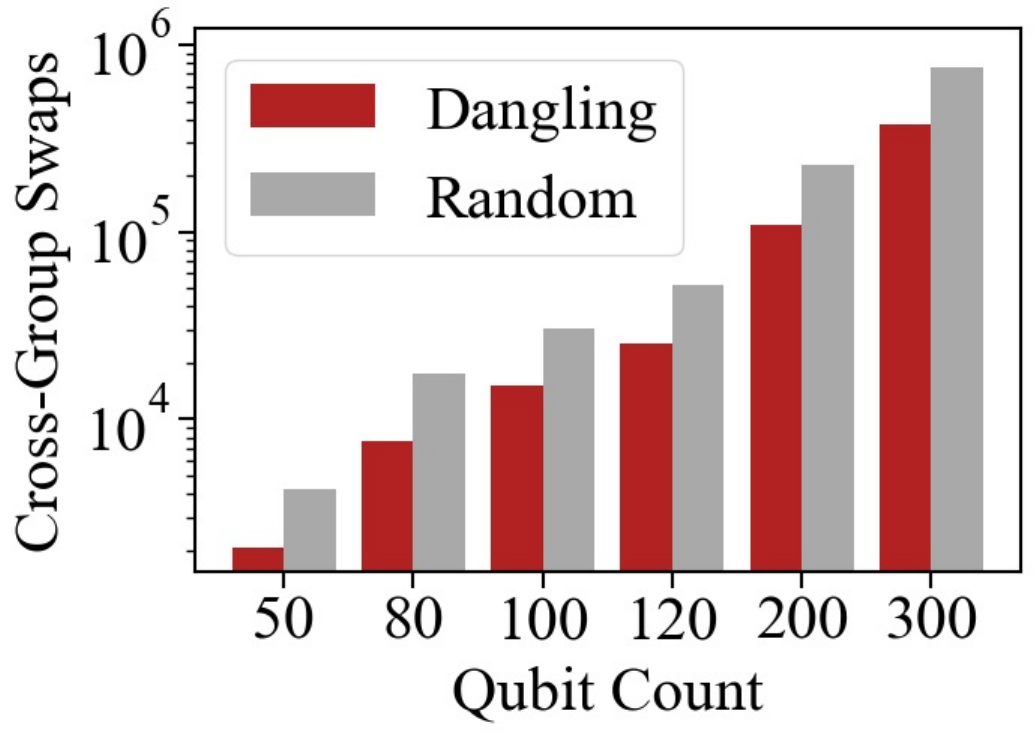}
         \caption{QFT Cross-group SWAP}
        \label{subfig:qftcross}
     \end{subfigure}
     \hfill  
     \begin{subfigure}[b]{0.22\textwidth}  
         \includegraphics[width=\linewidth]{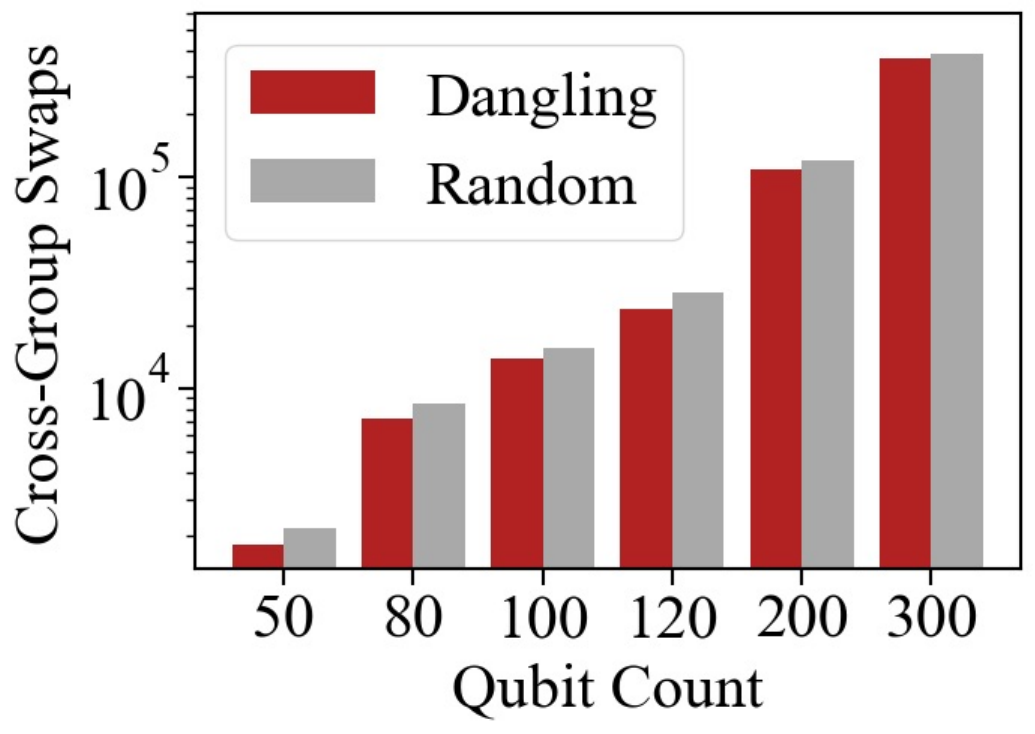}
         \caption{QAOA Cross-group SWAP}
        \label{subfig:qaoacross}
     \end{subfigure}
    \caption{Total SWAP Gates and Cross-Group SWAP Gates Comparison between Random and Dangling Qubit Placement.}
    \label{fig:swap_comparison}   
    \vspace{-3mm}
\end{figure}

As shown in Figure \ref{fig:random_vs_dangling_gate_count}, the dangling qubit method consistently reduces gate count across all qubit sizes, nearly halving the gate count in a 100-qubit circuit compared to the random method. Similarly, Figure \ref{fig:swap_comparison} shows a significant reduction in both total and cross-group SWAP gates, with the 100-qubit circuit seeing a drop from 15,117 to 7,216 cross-group SWAPs.

The considerable reduction in both gate count and SWAP gates highlights the superiority of the dangling qubit method, especially for larger circuits. 
This reduction is critical for minimizing overhead in distributed quantum systems, improving both scalability and circuit fidelity.

\subsection{Scalability and Robustness}
The DQC LNN method demonstrates excellent scalability, effectively reducing the exponential growth in gate count and SWAP operations as qubit numbers increase. By strategically placing non-local gates on dangling qubits, DQC LNN significantly lowers the total gate count and cross-group SWAP gates compared to random qubit placement. As the number of qubits grows, the difference between random and dangling placements becomes more pronounced; for instance, in a 300-qubit circuit, the dangling qubit method reduces the gate count to 1,673,560 compared to 3,249,886 with random placement. This suggests that DQC LNN will become even more advantageous for large-scale tasks involving thousands of qubits in the future.

\section{Conclusion and Discussion}
\label{section_conclusion}
In this paper, we proposed a distributed quantum compilation method based on the LNN architecture, utilizing dangling qubits for non-local gate operations. Our experiments demonstrated significant reductions in gate count and circuit depth, showcasing a scalable solution for DQC. The approach led to a 50\% reduction in gate count and minimized cross-group SWAP gates, resulting in more efficient quantum circuits and improved performance.

Our experiments with non-local gates are conducted on simulators, as currently the non-local gates in real quantum computers are still under implementation.
In real machines, non-local gates typically have lower fidelity and higher error rates, making cross-group SWAP gate count a even more important metric. Minimizing non-local interactions is key to improving fidelity, and future tests on actual hardware are needed to more accurately evaluate the impact.

Our current design uses a single pair of dangling qubits for non-local gates, while LNN can support more complex configurations with multiple pairs. This opens opportunities for further optimization by aligning dangling qubit connections with the needs of HLC-SFC algorithms like QFT. Pre-arranging qubits could reduce movements and SWAP operations, leading to an even more efficient compiler design. One limitation of our experiments lies in the use of a simulator environment, where the efficiency and fidelity of non-local gate construction were not fully considered. In real quantum machines, the distinction between local and non-local gates is crucial, as non-local gates generally exhibit lower fidelity and higher error rates. This is why we introduced the cross-group SWAP gate count as a key metric. Minimizing non-local interactions during compilation is essential for achieving higher fidelity and more robust quantum computations. Future experiments on actual quantum hardware will be necessary to fully explore the impact of this metric and the practical implications of our approach.

Our current implementation focuses on a straightforward chip-to-chip distributed architecture, using a single pair of dangling qubits for non-local gate operations. However, the LNN architecture has the potential for more complex configurations, where multiple pairs of dangling qubits could be employed simultaneously in distributed designs. This opens up opportunities to further optimize the compilation process by exploring how the topology of dangling qubit connections aligns with the specific demands of different quantum algorithms. For instance, in algorithms such as QFT, which involve highly structured problem graphs, qubits could be pre-arranged based on the algorithm's requirements. This tailored approach could result in even greater efficiency, as the qubits would be positioned to minimize unnecessary movements and SWAP operations, leading to an optimized compiler design that aligns with both the hardware topology and the algorithm’s structure.



\clearpage
\bibliographystyle{ieeetr}
\bibliography{references}

\end{document}